\documentclass[epj,twocolumn]{webofc}
\usepackage[varg]{txfonts}   
\usepackage{amssymb}
\usepackage{url}
\def\66a{3C~66A}
\def\ver{VER~J0521+211}
\def\lat{{\em Fermi}-LAT}
\def\wcom{W~Comae}
\def\pks{PKS~1424+240}
\def\s5{S5~0716+714}
\def\bl{BL~Lacertae}
\def\b2{B2~1215+30}
\def\apj{ApJ}
\def\apjl{ApJL}
\def\aap{A\&A}
\def\nat{Nature}
\def\mnras{MNRAS}

\woctitle{The Innermost Regions of Relativistic Jets and Their Magnetic Fields}
\begin{document}
\title{Observations of low- and intermediate-frequency-peaked BL Lacs
above 100 GeV with VERITAS}

\author{M.~Errando\inst{1}\fnsep\thanks{\email{errando@astro.columbia.edu}} 
        for the VERITAS Collaboration
}

\institute{Department of Physics and Astronomy, Barnard College, Columbia University, NY 
10027, USA}

\abstract{%
Most of the $\sim50$ blazars detected to date at TeV energies ($E>0.1$\,TeV) are 
high-frequency-peaked BL Lacs (HBLs). Only a handful episodic TeV detections of low- and 
intermediate-frequency-peaked BL Lacs (LBL/IBLs, with synchrotron peak frequencies in the infrared 
and optical regime) have been reported, typically during high-flux states. 
The VERITAS array, a ground-based TeV observatory located in southern Arizona has observed five known TeV LBL/IBLs since 2009: 
\66a, \wcom, \pks, \s5\ and \bl, with at least 5-10 hours/year, which so far resulted in the 
detection of a bright, sub-hour timescale gamma-ray flare of \bl\ in June 2011. We also report the 
detection and characterization of two new IBLs: \ver\ and \b2.}
\maketitle
\section{Introduction}
\label{intro}

The population of TeV blazars\footnote[2]{Detected at $E>0.1$\,TeV, which is the typical energy 
range of current imaging atmpospheric Cherenkov telescopes.} is dominated by high-frequency-peaked 
BL Lacs (HBLs), comprising 75\% of the total TeV blazar count \cite[41 out of 55,][]{tevcat}. HBLs 
show a maximum of their synchrotron emission in the X-ray regime, and display a low Compton 
dominance $R_c=L_{HE}/L_{sy}\lesssim1$, defined as the ratio between the luminosity of the 
high-energy component of the spectral energy distribution and that of the synchrotron emission. At 
parsec scales, radio imaging of TeV 
HBLs shows relatively weak jets with only stationary or sub-luminal components \cite{piner}.

\begin{figure}[t]
\centering
\includegraphics[width=0.95\columnwidth,clip]{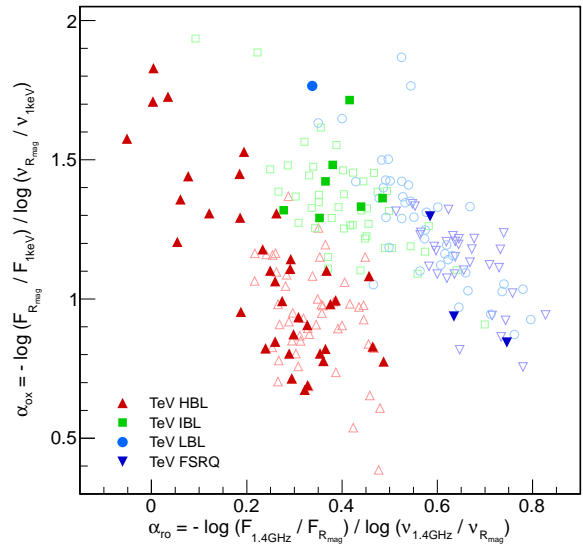}
\caption{Spectral slopes of the synchrotron component for gamma-ray blazars.
The effective spectral indices $\alpha_{\mathrm{ro}}$ and $\alpha_{\mathrm{ox}}$ are defined in the
usual way between 1.4\,GHz, 6590\,\AA\ and 1\,keV. Adapted from \cite{0521}.}
\label{fig:slopes}       
\end{figure}
\begin{table}[b]
\begin{footnotesize}
\centering
\begin{tabular}{lllll}
\hline
\hline
 & RA & Dec& $z$  & ref. \\\hline
\66a & 02 22 41	& +43 02 35 & 0.41? & \cite{66a,magic-66a}\\
\ver & 05 21 55	 & +21 11 24 & 0.108 & \cite{0521}\\
\s5 & 07 21 53 & +71 20 36	& 0.31? & \cite{magic-s5}\\
\b2 & 12 17 52 & +30 07 01	& 0.13?	& \cite{magic-1215,1215}\\
\wcom & 12 21 31	& +28 13 59 & 0.102 & \cite{wcom,wcom-flare}\\
\pks & 14 27 00 & +23 47 40 & $>0.6$ & \cite{1424}\\
AP~Librae &15 17 41&-24 22 19	&0.049 & \cite{hess-aplib}\\
MAGIC J2001+435 & 20 01 13 & +43 53 02& ? & \cite{magic-2001}\\
\bl & 22 02 43 & +42 16 40 & 0.069 & \cite{magic-bllac,bllac-flare}\\
\hline
\hline
\end{tabular}
\end{footnotesize}
\caption{TeV-detected IBLs and LBLs.}
\label{tab:sources}       
\end{table}

In recent years, the improved sensitivity of the current generation of ground-based Cherenkov 
telescopes (VERITAS, MAGIC, and H.E.S.S.) revealed TeV emission from a handful of low- and 
intermediate-frequency-peaked BL Lacs (LBLs/IBLs), listed in Table.~\ref{tab:sources} \cite{magic-bllac,wcom,66a,magic-s5,hess-aplib,1424,magic-2001,magic-1215, 0521}. 
LBLs and IBLs show different spectral properties than HBLs, with synchrotron emission peaking in the 
infrared/optical regime and falling towards the X-ray band (Fig.~\ref{fig:slopes}). In general, 
LBLs and IBLs are more
powerful, more luminous, and have a richer jet environment compared to HBLs. 
The accretion rate is thought to be the main driver of these differences \cite{meyer,giommi}.  
In this scheme, LBLs
accrete close to the Eddington rate, producing powerful and 
luminous jets, and an optically thick accretion disk whose radiation
ionizes the broad line region (BLR) and is reflected by the dust torus, generating sources of 
external low-energy photons for Compton scattering. On the other hand, HBLs accrete at highly 
sub-Eddington rates, in a radiatively inefficient regime, producing weaker jets that propagate in a 
``cleaner'' environment. IBLs may represent an intermediate case in terms of their accretion 
properties, although jet alignment has also been suggested as an explanation for their lower peak 
frequencies \cite{meyer}.
In the lowest frequency band, a recent 
compilation of 15\,GHz radio images \cite{lister} found superluminal components in \66a, \s5, 
\wcom, AP~Librae, and \bl. For comparison, only one TeV HBL showed superluminal motion in the same 
study.

These proceedings report on observations of LBLs and IBLs with VERITAS in the gamma ray band 
 ($E>0.1$\,TeV), focusing on the recent detections of two new intermediate BL~Lacs (\ver\ and \b2) 
and a bright TeV flare from \bl. More details on these results have been provided in references 
\cite{0521,1215,bllac-flare}. The gamma-ray flux of the Crab Nebula is often referred as a 
reference unit throughout the text, being $1\,\mathrm{Crab} = 2.1 \times 
10^{-10}\mathrm{cm^{-2}}\mathrm{s^{-1}}$ above 0.2\,TeV \cite{hillas}.

\section{Discovery of TeV emission from \ver}
\label{ver}
\ver\ was suggested as a TeV candidate during a search for clusters of $>30$\,GeV photons in the 
\lat\ data. Observations with VERITAS in October 2009 resulted in a significant detection at 
energies above 0.2\,TeV in less than four hours of observation \cite{0521}, implying a relatively 
bright TeV flux at a level of $F_{>0.2\,\mathrm{TeV}} \sim 0.1$\,Crab. Follow-up observations 
revealed a bright flare reaching $\sim 0.3$\,Crab with a flux doubling time of $\sim 1$\,day. 
Incidentally, \ver\ lies only $3^{\circ}$ away from the Crab Nebula, the first-detected TeV 
emitter and the brightest and best-studied gamma-ray 
source in the sky. The Whipple 10m telescope, and later the VERITAS array, had been observing the 
Crab Nebula and the region around it for more than two decades, just missing \ver\ outside 
the edge of their field of view.

\begin{figure}[t]
\centering
\includegraphics[width=0.95\columnwidth]{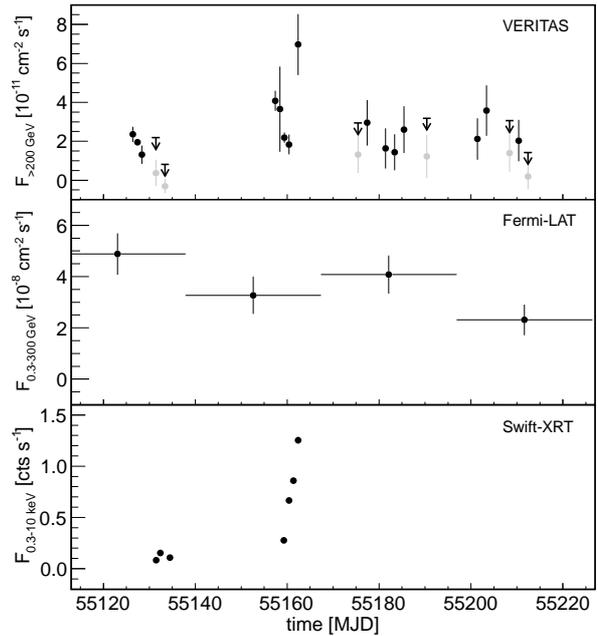} 
\caption{VERITAS ($E>0.2$\,{TeV}), \lat\ ($0.1-300$\,GeV), and {\it Swift}-XRT (0.3-10\,keV) light
curves of \ver. Photon fluxes are calculated in 1-day bins for VERITAS and {\it Swift}-XRT,
and 29.5 days for {\it Fermi}-LAT. The VERITAS light curve shows significant flux points (black
dots) when the signal exceeds $2\sigma$, and 95\% confidence level upper limits (black arrows) 
together with flux points (gray dots) otherwise. The error bars on the {\it Swift}-XRT 
rates are at the $\sim 2-7\%$ level, and are not visible in the plot. Taken from \cite{0521}.}
\label{fig:ver}       
\end{figure}

The TeV emission detected with VERITAS was spatially associated with RGB~J0521.8+2112, an 
unidentified radio and X-ray source. A complete multiwavelength campaign 
triggered after the VERITAS detection unambiguously classified \ver\ as a BL~Lac-type blazar. The 
evidence included a one-sided jet resolved in 15\,GHz VLBA images, polarized optical emission;
point-like, variable X-ray and gamma-ray emission (Fig.~\ref{fig:ver}), and a featureless optical 
spectrum. Later observations revealed optical absorption lines, indicating a redshift of $z=0.108$ 
\cite{shaw}. Radio imaging also shows electric polarization vectors perpendicular to the jet 
ridgeline, indicative of optically thin synchrotron emission associated with a well-ordered 
magnetic 
field aligned in the direction of the jet axis.

The broadband synchrotron emission spectrum from \ver\ shows a peak in the optical band, suggesting 
a classification as an IBL. However, during the X-ray and TeV flare in MJD~55162 
(Fig.~\ref{fig:ver}) its synchrotron properties resembled those of TeV HBLs, mainly due to a 
hardening of the X-ray spectrum during the high-flux state. Even if it did not figure in 
TeV-candidate catalogs before the launch of {\it Fermi} due to its low Galactic latitude, \ver\ 
ranks among the brightest blazars known at TeV energies. At a distance of $z=0.108$, the 
gamma-ray luminosity of \ver\ exceeds $2 \times 10^{44}\mathrm{erg}\,\mathrm{s}^{-1}$, 
significantly brighter than classical northern blazars such as Mrk~421, Mrk~501 and 1ES~1959+650. 
Future multiwavelength observations of \ver\ will help to extend the detailed spectral modeling 
available for nearby HBLs to a more luminous non-HBL blazar.

\section{TeV detection and characterization of \b2}
\label{b2}

\begin{figure}[t]
\centering
\includegraphics[width=0.90\columnwidth]{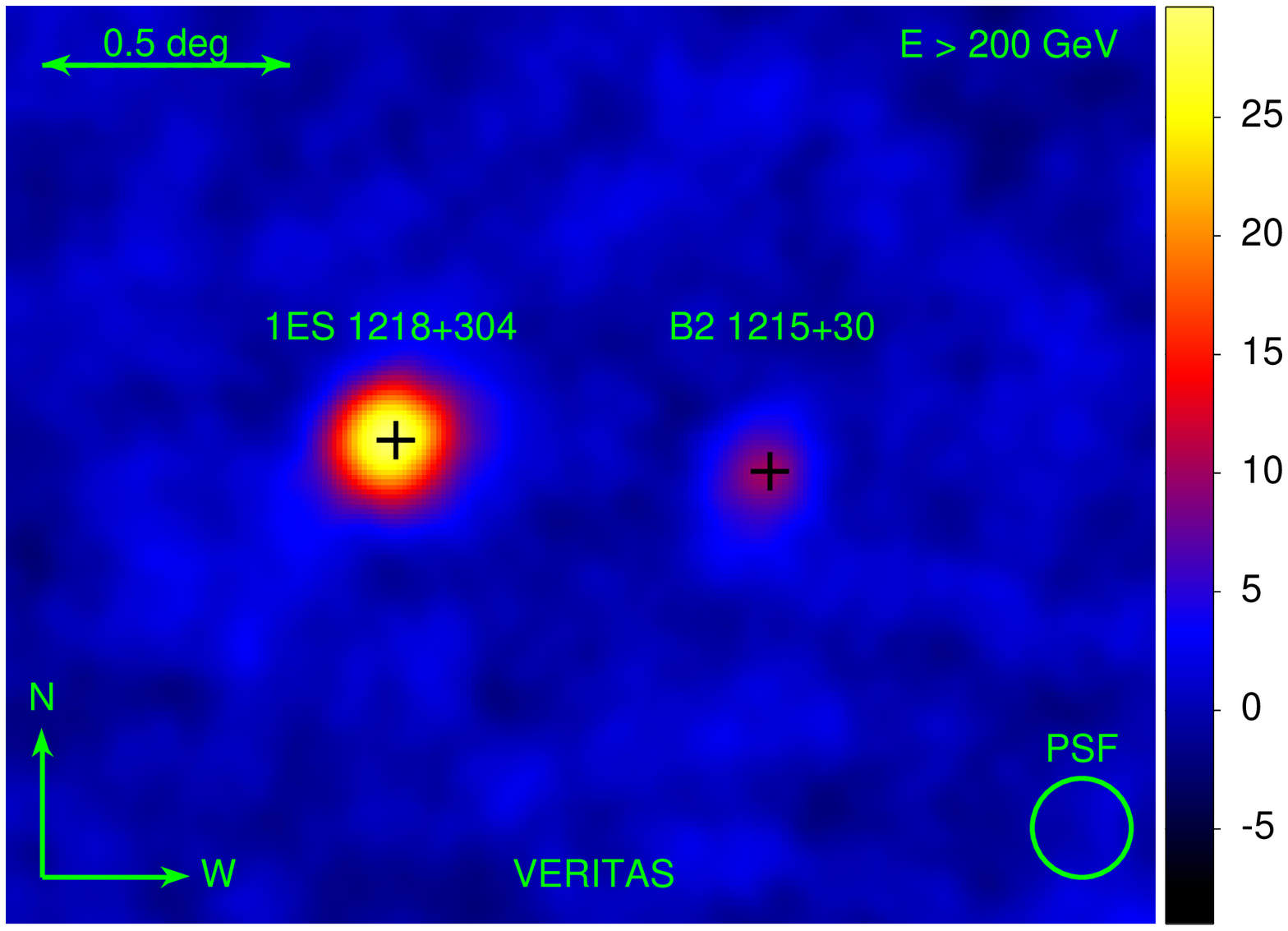} 
\includegraphics[width=0.90\columnwidth]{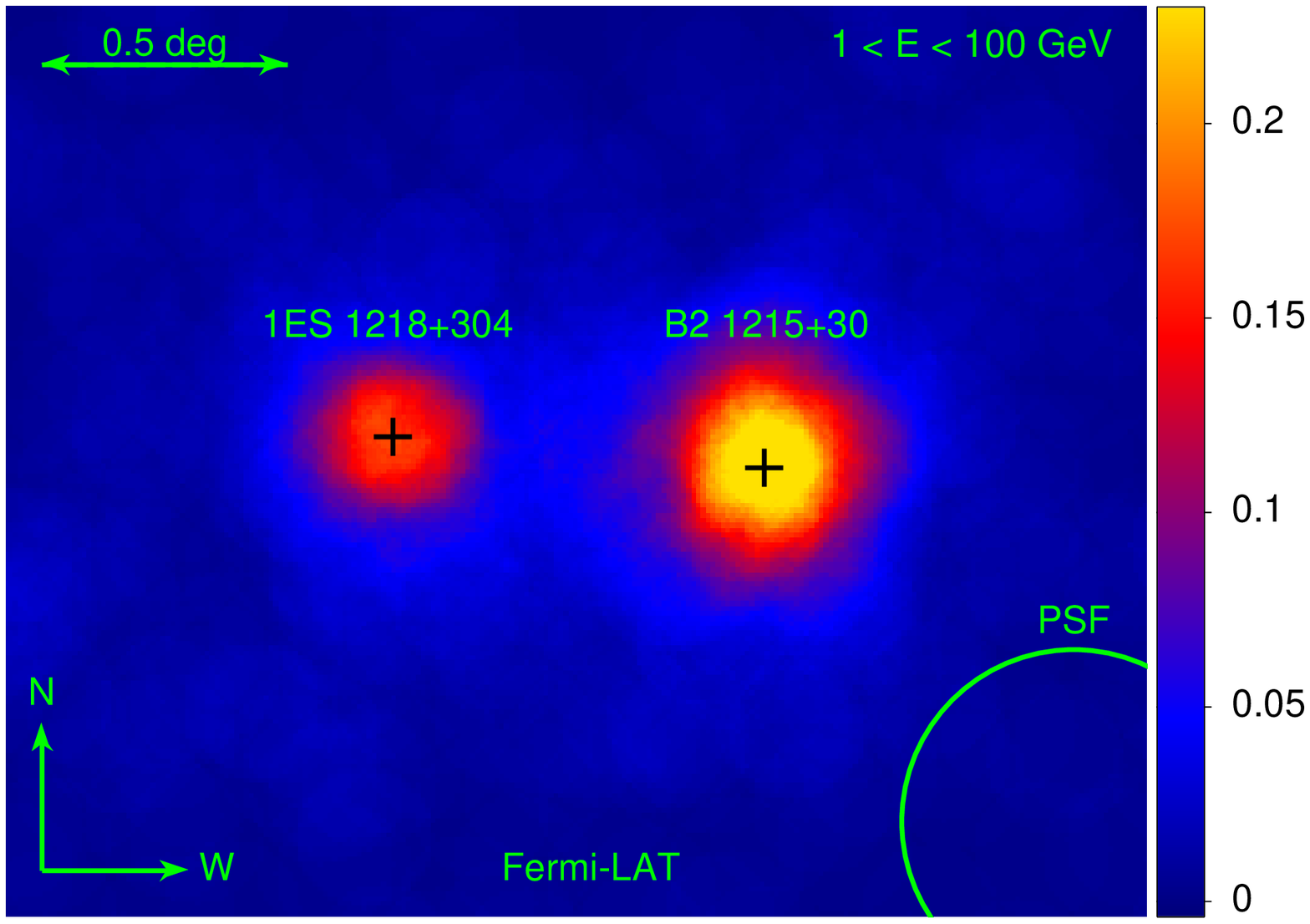} 
\caption{Sky map of the gamma-ray emission measured towards the direction of \b2\ 
in two different bands. The top panel (adapted from \cite{1215}) shows the 
significance of the gamma-ray signal above 0.2\,TeV as measured with VERITAS in 2011. The bottom 
panel shows photon counts selecting front-converting photons with $E>1$\,GeV in the \lat\ data (Aug 
2008 - Sep 2013). The green circles indicates the 68\% containment angle of both detectors, and the 
radio locations of  \b2\ and 1ES~1218+304 are marked 
with black crosses.}
\label{fig:1215}       
\end{figure}

\b2\ is one of the five sources that were used to define the class of BL~Lac-type objects, 
together with OJ~287, \wcom, AP~Librae, and \bl\ \cite{strittmatter}. Its distance is uncertain, 
although spectroscopic redshifts of $z=0.130$ and $z=0.237$ can be found in the literature.

TeV emission from \b2\ was reported by MAGIC after observations triggered during an optical high 
state \cite{magic-1215}. The source is only $0.76^\circ$ away from 1ES~1218+304, a well-studied 
TeV HBL that has been extensively monitored with VERITAS since 2009. 
VERITAS collected a gamma-ray signal from \b2 at $8.9\sigma$ level after 82\,hr of 
exposure-corrected observations (most of them 
targeting 1ES~1218+304) spanning over three yearly observing seasons. 1ES~1218+304 is a bright HBL 
($F_{>0.2\,\mathrm{TeV}} \sim 0.06$\,Crab) showing a power-law spectrum in the TeV band with 
$\Gamma_{1218+30} = 3.1 \pm 0.3$ \cite{1218}, while the measured spectral index for \b2\ is 
significantly softer ($\Gamma_{1215+30} = 3.6 \pm 0.4$). The spectral differences can be visualized 
in Figure~\ref{fig:1215}, showing how \lat\ and ground-based gamma-ray telescopes have different 
sensitivities to distinct source classes. \b2, with $F_{>0.2\,\mathrm{TeV}} \sim 0.03$\,Crab, is 
significantly weaker than 1ES~1218+304 in the TeV band, but the picture reverses at lower energies, 
where \lat\ measures a much brighter photon flux from \b2.

The gamma-ray signal from \b2\ recorded by VERITAS shows clear flux variability on time scales of 
several months, with a prolonged high state in 2011, in agreement with \cite{magic-1215}. 
Significant variability on shorter time scales could not be resolved. Observations in the optical 
band (Super-LOTIS, MDM, {\it Swift}-UVOT), 
X-ray ({\it Swift}-XRT), and gamma rays (\lat) quasi-simultaneous with the 2011 VERITAS detection 
were used to construct a complete spectral energy 
distribution, displaying a synchrotron peak in the ultraviolet regime and therefore 
confirming \b2\ as an IBL. The multiwavelength data were successfully described 
with a synchrotron self-Compton model with parameters similar to those of other TeV-detected 
blazars.

\section{TeV flare of \bl}
\label{bl}

\begin{figure*}[t]
\centering
\includegraphics[height=2.57in]{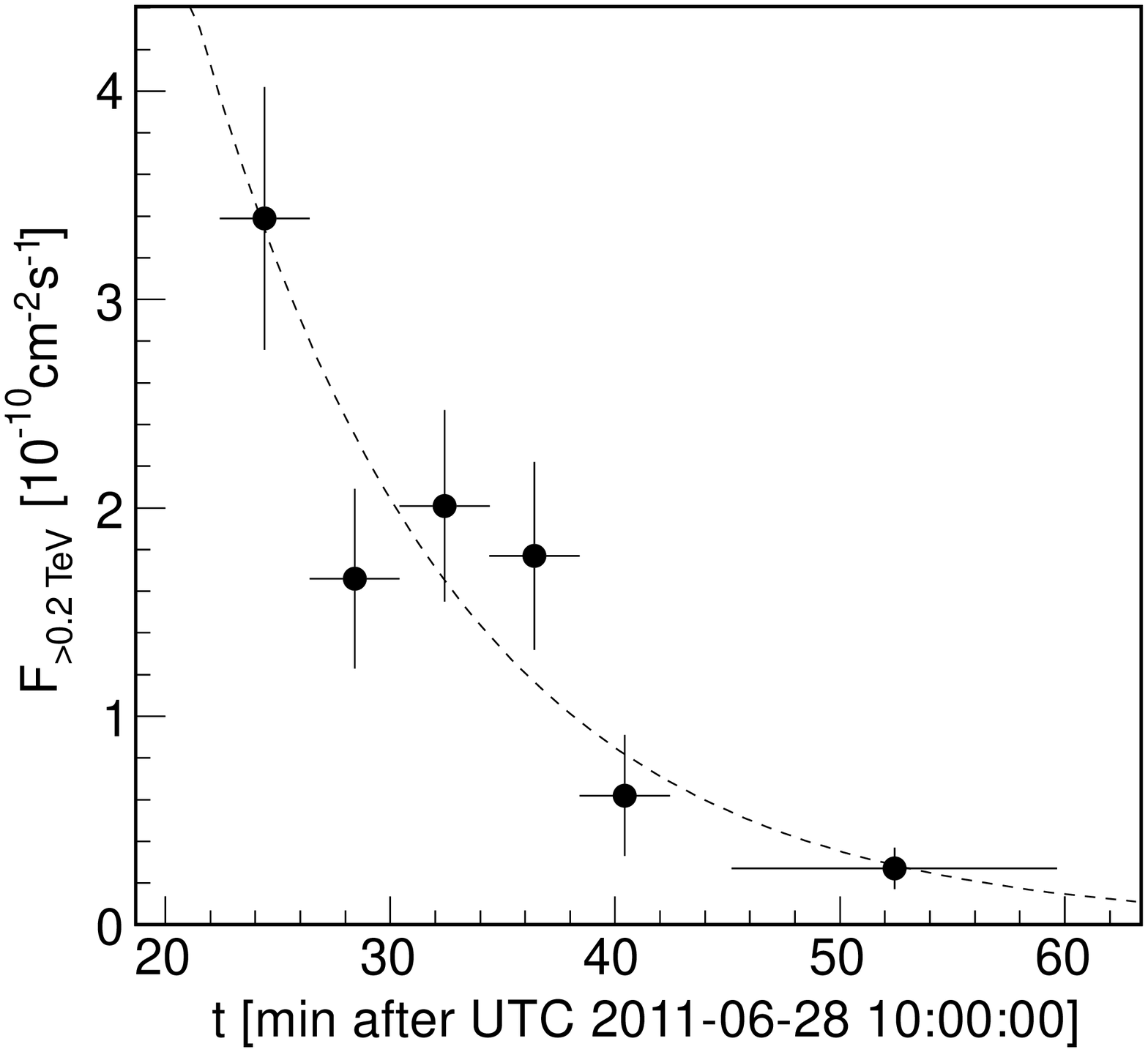} 
\hspace{1cm}
\includegraphics[height=2.53in]{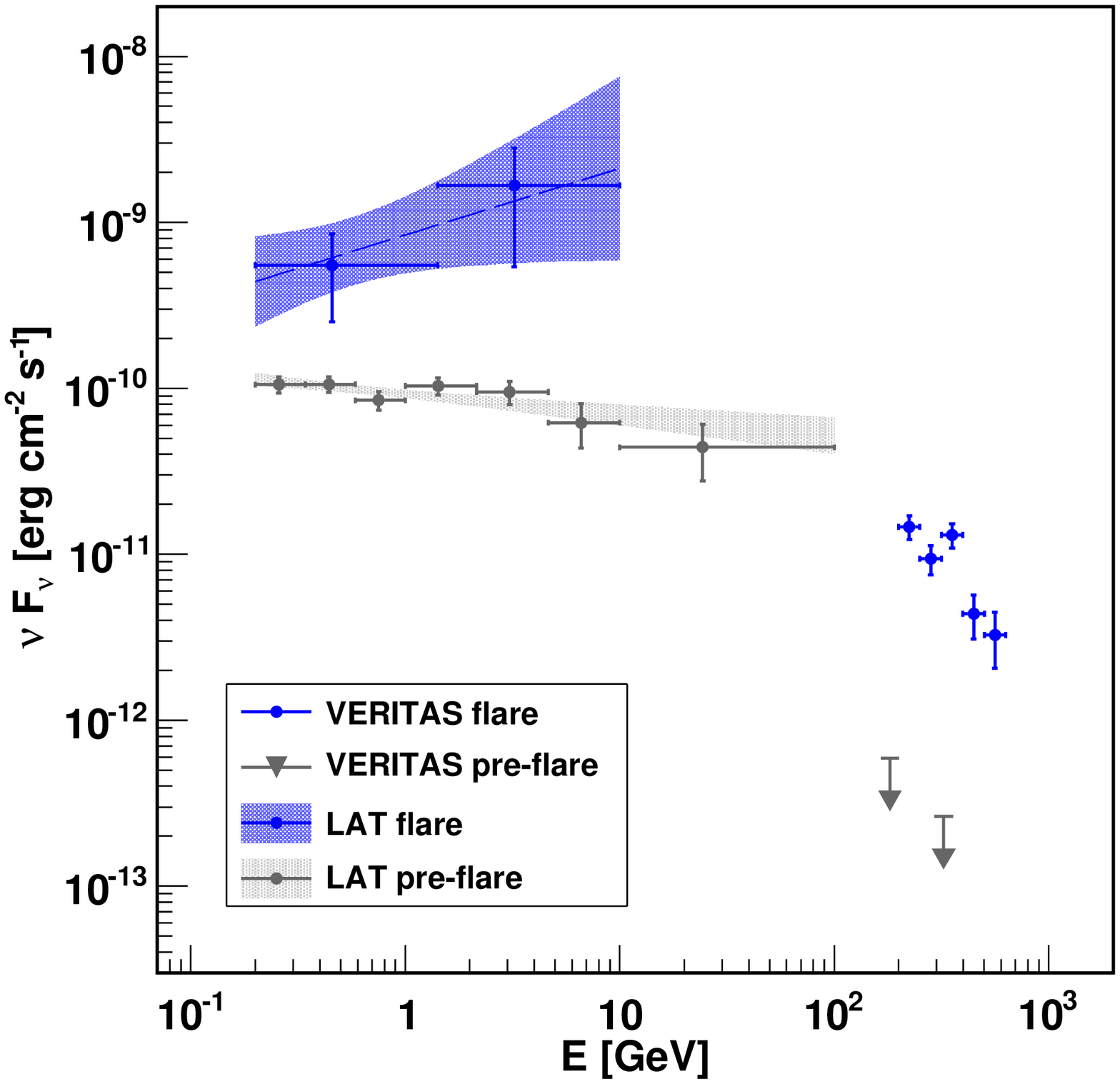} 
\caption{(left) Gamma-ray flux from \bl\ measured by VERITAS on 2011 June 28. Flux is binned in 
intervals as short as 4\,min, and the dashed line indicates the best-fit exponential flux 
decay yielding to a $t_{var}=13 \pm 4$\,min. (right) Broadband gamma-ray spectrum of \bl\  
from simultaneous \lat\ and VERITAS observations on 2011 June 28. Figures adapted from 
\cite{bllac-flare}. }
\label{fig:bllac}  
\end{figure*}

The eponymous blazar \bl\ was detected by VERITAS during a short-lived TeV flare on 2011 June 28 
\cite{bllac-flare}. The gamma-ray flux over the 35\,min of observations was $F_{>0.2\,\mathrm{TeV}} 
\sim 0.3$\,Crab, ten times brighter than in the original TeV discovery \cite{magic-bllac}. 
Figure~\ref{fig:bllac} shows the measured light curve resolved down to 
4-min time bins, where intra-night variability is first observed for this object in the gamma-ray 
band. The measured flux has a decay time $t_{var} = 13 \pm 4$\,min, and the peak flux reaches 
$\sim1.6$\,Crab. 

Causality implies that the observed variability timescale is related to the size of the gamma-ray emission region by 
$R \lesssim c\, t_{var}\, \delta \, (1+z)^{-1}  \sim (t_{var}/13\,\mathrm{min})\, (\delta/10) \cdot 
7\times10^{-5}$\,pc, where $\delta$ is the Doppler factor of the emitting plasma. Correlation of the 
gamma-ray flaring event with the emergence of a superluminal knot in radio images (also reported in 
the discovery TeV emission from \bl\ in 2005-6 \cite{marscher}) has been interpreted as indication 
of the gamma-ray dissipation region to be at parsec scales from the central black hole. Under that 
scenario, the compact size of the emitting region of $R\lesssim 7\times10^{-5}$\,pc would represent 
only a small fraction of the jet cross section at parsec scales, which would be $\sim 0.1$\,pc 
assuming a conical collimated jet. That would suggest that a small emitting region downstream of 
the jet can be responsible for most of its radiative output.

The broadband gamma-ray spectrum during the flare shows a sharp break between 
$\Gamma_{\mathrm{GeV}}=1.6\pm0.4$ measured by \lat\ and $\Gamma_{\mathrm{TeV}}=3.8\pm0.3$ seen by 
VERITAS (Fig.~\ref{fig:bllac}). A spectral break at $\sim 15$\,GeV would be expected due to 
Klein-Nishina suppression of the electron-photon cross section if the dominant component 
of the gamma-ray emission is inverse-Compton scattering of external photons from the BLR 
\cite{hpblazars}, as indicated in previous multiwavelength studies of 
the source both in quiescent and flaring states \cite{madejski,fermi-bllac}. Additional steepening 
of the gamma-ray spectrum could be due to pair production by gamma-rays on a dense photon field 
with frequencies similar to those of BLR photons \cite{poutanen}.

\vspace{0.2cm}
A coherent picture is therefore difficult to put forth, as BLR photons are not expected to be 
dominant at parsec scales but closer to the central black hole. Further observations of \bl, and of 
TeV LBLs and IBLs in general are needed to elucidate the location of the gamma-ray emitting region 
and the role of external photons in the gamma-ray production and absorption, as well as how the 
energy dissipation processes change between quiescent states and bright flaring events.

\vspace{0.4cm}
\begin{small}
VERITAS is supported by grants from the U.S. Department of Energy Office of Science, the U.S.
National Science Foundation and the Smithsonian Institution, by NSERC in Canada, by Science
Foundation Ireland (SFI 10/RFP/AST2748) and by STFC in the U.K. We acknowledge the excellent work of
the technical support staff at the Fred Lawrence Whipple Observatory and at the collaborating
institutions in the construction and operation of the instrument.
\end{small}

\end{document}